# INTENSIVE NEUTRINO SOURCE ON THE BASE OF LITHIUM CONVERTER

V. I. Lyashuk [1,2] and Yu. S. Lutostansky [2]

[1] *Institute for Nuclear Research, Russian Academy of Science, Moscow, Russia*
[2] *National Research Center "Kurchatov Institute", Moscow, Russia*

**Abstract:**

An intensive antineutrino source with a hard spectrum ( $E_{\tilde{\nu}}^{max}$ = 13 MeV and $<E_{\bar{\nu}}>$ = 6.5 MeV) can be realized on the base of β⁻-decay of short living isotope $^8$Li ( $T_{1/2}$ = 0.84 s). The $^8$Li isotope (generated in (n,γ)-activation of $^7$Li isotope) is a prime perspective antineutrino source owing to the hard $\tilde{\nu}_e$ –spectrum and square dependence of neutrino cross section on the energy ( $\sigma_\nu \sim E_\nu^2$ ). Up today nuclear reactors are the most intensive neutrino sources. Antineutrino reactor spectra are formed by $^{235}$U, $^{238}$U, $^{239}$Pu and $^{241}$Pu isotopes which cause large uncertainties in the summary antineutrino spectrum at $E_{\bar{\nu}}$ > 6 MeV. Use of $^8$Li isotope allows to decrease sharply the uncertainties or to exclude it completely. An intensive neutron fluxes are requested for rapid generation of $^8$Li isotope. The installations on the base of nuclear reactors (of steady-state or pulse neutron fluxes) can be an alternative for nuclear reactors as "traditional" neutron sources. It is possible creation of neutrino sources another in principle: on the base of beam-dumps of large accelerators plus $^7$Li converter; on the base of tandem of accelerators, neutron generating targets and lithium converter. An intensive neutron flux (i.e., powerful neutron source) is requested for realization of considered neutrino sources (neutrino factories). Different realizations of lithium antineutrino sources (lithium converter on the base of high purified $^7$Li isotope) are discussed: static regime (i.e., without transport of $^8$Li isotope to the neutrino detector); dynamic regime (transport of $^8$Li isotope to the remote detector in a closed cycle); an operation of lithium converter in tandem of accelerator with a neutron-producing target on the base of tungsten, lead or bismuth. Different chemical compounds of lithium (as the substance of the converter) are considered. Heavy water solution of LiOD is proposed as a serious alternative to high-pure $^7$Li in a metallic state.

## 1. Introduction. Advantages and disadvantages of the reactor $\tilde{\nu}_e$ -spectrum

The antineutrino spectrum of nuclear reactors is formed by β⁻-decay of nuclear reactor fission fragments. The spectrum is sharply decreasing and has an energy $E_{\bar{\nu}}$ ≤ 10 MeV. Cross sections of interactions of reactor antineutrino with proton, electron and deuteron are exclusively small – in the interval $10^{-46}$ –$10^{-43}$ cm$^2$/fission. In fact the full $\tilde{\nu}_e$ –flux of the reactor (99.8% for water-moderated reactor types) is ensured by four isotopes — $^{235}$U, $^{239}$Pu, $^{238}$U, $^{241}$Pu [1 - 3]. An experimental equilibrium $\tilde{\nu}_e$ –spectra of nucleus-fission products of these four isotopes (the spectra are normalized per one fission) are presented in the fig. 1 [4 - 6]. An experimental data of $\tilde{\nu}_e$ –spectrum for $^{238}$U isotope was published only in 2014. The yield of $^{238}$U to the summary neutrino output for water-moderated reactor types (according to FRM-II reactor in Garching, Germany) is evaluated as 10% [6]. As clear from fig.1 the four spectra are dropping rapidly as energy increases (that is especially negatively for registration of threshold reactions): at increase $E_{\bar{\nu}}$ from 2 MeV to 4, 6 and 8 MeV the $^{235}$U-spectrum drops in 5, 35 and 956 times, respectively.

One more complication is dependence of partial spectra from nuclear fuel composition which vary in time in operation period and in case of reactor stops. The isotope $^{235}$U burns away, but the contributions of $^{239}$Pu, $^{238}$U, $^{241}$Pu rise (see fig.2) [2, 7].

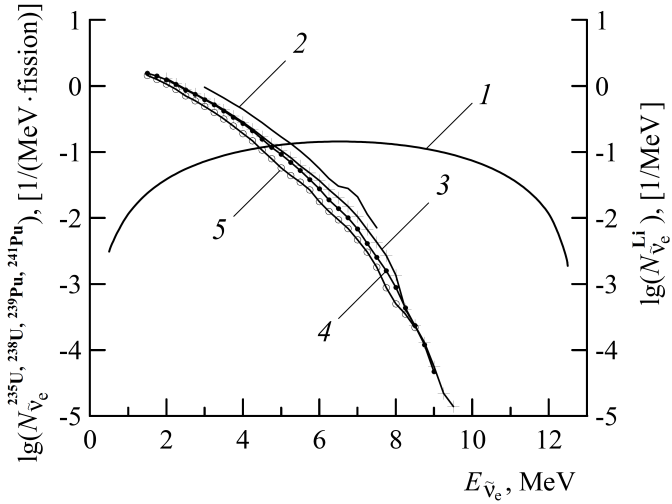

Fig. 1. $\tilde{\nu}_e$ - spectra from $^{235}$U, $^{238}$U, $^{241}$Pu, $^{239}$Pu and $\beta^-$-decay of $^8$Li: 1 - $^8$Li; 2 – $^{238}$U; 3 – $^{235}$U; 4 – $^{241}$Pu; 5 – $^{239}$Pu.

$$N_{\tilde{\nu}}(t,E) = \sum_{i=5,8,9,1} a^i(t) N_{\tilde{\nu}}^i(E),$$

The changing in composition leads to variation in $\tilde{\nu}_e$–fluxes which are recalculated by means of correction factors for four isotopes [8, 9]. The summary neutrino flux can be written as additive function of fluxes: from fission fragments; from beta decay of heavy (transuranium) nuclei (produced in (n,γ)- and (n,2n)-reactions); and β-decay of (n,γ)-activated elements of the constructions and water. The flux from fission fragments from $^{235}$U, $^{239}$Pu, $^{238}$U and $^{241}$Pu can be presented as summation with factors $a^i(t)$ which depend on time from the beginning of the reactor life-time:

where $i$ = 5, 8, 9, 1 are mean sum for isotopes. So from the beginning of the reactor life-time (~290 to 330 days) the total $\tilde{\nu}_e$ –flux increases, for example after 10 and 330 days the fluxes are 6.39 and 6.78 $\tilde{\nu}_e$/fission respectively [10].

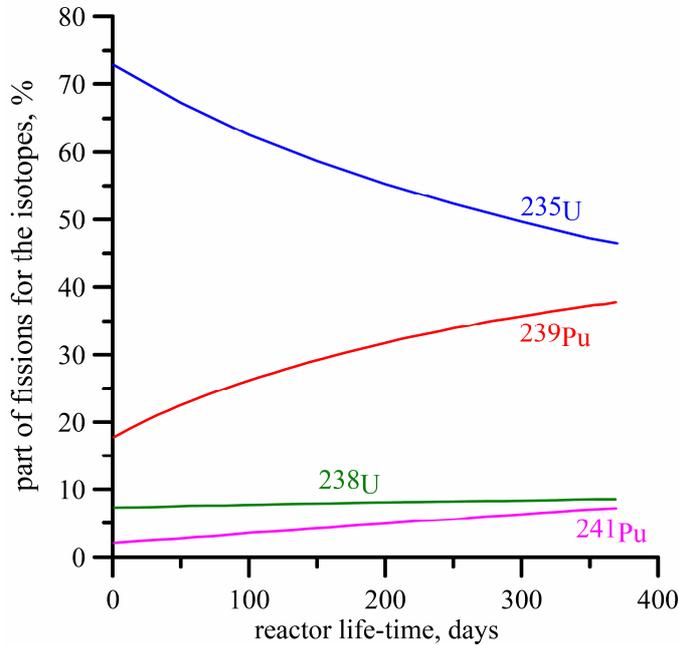

Fig.2. Evolution for number of fissions from $^{235}$U $^{239}$Pu, $^{238}$U and $^{241}$Pu (four main isotopes in nuclear fuel) during the reactor life-time (2nd and 3rd reactor life-time) for water-moderated reactor types [7].

During the reactor life time (as nuclear fuel composition varies) the mean energy $E_f$ (released per one fission) increases on 1.5% [9] It is important as the summary neutrino flux depends on number of fissions in the active zone as: $n_f(t)=W(t)/E_f(t)$, where $W(t)$ is a full reactor heat power for the current time. It is considered that for high pressure water reactor the current value of full heat power can be established with accuracy about 2% [11].

An additional unaccounted errors for $\tilde{\nu}_e$ –flux evaluation appear during reactor stops due to permanent presence of cooling pond for spent fuel. These errors can rise up to 1%. It is usually considered that partial neutrino spectra $N_{\tilde{\nu}}^i$ reach equilibrium conditions after 1 day from the reactor life time. But this evaluation must be corrected too as the summary spectrum increases from the beginning to the end of the reactor life time on the value (5-6)% [10].

An experimental $\tilde{v}_e$–spectrum of β−–decay of nuclei-fission fragments (for $^{235}$U, $^{239}$Pu, $^{238}$U, $^{241}$Pu) are recovered from β−–spectra of these isotopes. The direct registration of β−–spectra by electrons is possible only for part of decay chains: the other chains are identified by means of γ–quantum. In case of large branching the reproduction of channel probabilities, unknown decay schemes and the final products by means of γ–spectroscopy becomes problematic. The model $\tilde{v}_e$–spectrum calculations have been done for solve of the problem [12-14] including: parametrization of the effective charges for fragments [12]; an influence of corrections (radiation and coulomb ones, weak magnetism) [13]; yields of forbidden transfers [14].

In the model calculation of $\tilde{v}_e$–spectrum [12] (the 550 isotopes and 8000 chains were included) the cumulative $\tilde{v}_e$–fluxes (per fission) from $^{235}$U, $^{239}$Pu, $^{241}$Pu were corrected on 2.4, 2.9 and 3.2% respectively refer to [4, 5, 15]. The close results were obtained in the works [13] (model for 845 nuclei and 10000 chains): the corrections for $^{235}$U, $^{239}$Pu and $^{241}$Pu were 3.1, 3.1 and 3.7% refer to [4, 5]. The work [14] (which include 1500 forbidden transfers) also indicated on similar systematic errors. It was concluded the reactor $\tilde{v}_e$–spectrum is known with averaged accuracy less than ~4% and the level of accuracy can be caused by data on ~(25-30)% of forbidden transfers [14].

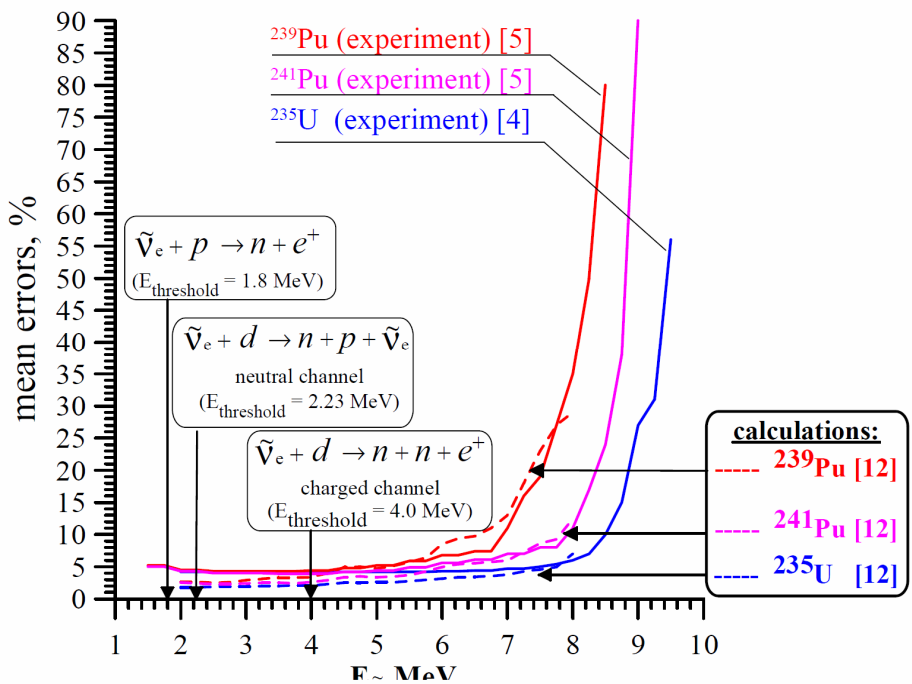

Fig.3. Errors for the obtained antineutrino spectra (from three isotopes of nuclear fuel - $^{235}$U $^{239}$Pu, and $^{241}$Pu) in dependence on the energy. Solid line – experimental results from ILL (Institute Laue Langevin) for $^{235}$U [4], $^{239}$Pu [5], and $^{241}$Pu [5]; dotted line – model calculation in the work [12].

The errors in reactor $\tilde{v}_e$–spectrum increase very significantly beginning from the energy about 6 MeV. For more higher energies the situation looks dramatically. So the errors rise up to: 56% at 9.5 MeV for $^{235}$U; 90% at 9.0 MeV for $^{241}$Pu; 80% at 8.5 MeV for $^{239}$Pu; 30% at 7.5 MeV for $^{238}$U [4-6]. The experimental $^{238}$U-spectrum reveal the 10%-discrepancy in comparison with cumulative $\tilde{v}_e$–spectra which obtained in calculations [6]. The dependence of errors for $\tilde{v}_e$–spectrum for experimental data and model calculations for $^{235}$U, $^{239}$Pu and $^{241}$Pu is shown in the fig. 3 [4, 5, 12]. Here the most significant deviations (starting from ~4%) of model data to experimental ones appear for $^{239}$Pu.

As a results the significant uncertainties in antineutrino nuclear reactor spectra and unaccounted addition to the summary $\tilde{\nu}_e$–spectrum (up to 6% in total flux; see works [10, 12 – 14]) make exclusively complicated the interpretation of neutrino oscillation experiments.

Significant errors in $\tilde{\nu}_e$–reactor spectrum lead to large errors in registration of threshold reactions of $(\tilde{\nu}_e, p)$ and $(\tilde{\nu}_e, d)$. It is especially important for the charged channel of $(\tilde{\nu}_e, d)$ - interaction taking in mind the dependence ($\sigma_\nu \sim E_\nu^2$) and the fact that at the energy $E > 4.0$ Mev (i.e., $E_{\text{threshold}}$ for $(\tilde{\nu}_e, d)$ -charged channel) the $\tilde{\nu}_e$–spectrum drops significantly (see fig. 1).

The nuclear reactors and isotope sources are the most widely used as intensive neutrino sources. The isotope sources have some advantages: the known characteristics, usability and availability. But isotope sources yield to reactors in fluxes [3]. The serious disadvantage is significant decrease of intensity of artificial neutrino sources in time. As example the neutrino flux of the unique source of the base of $^{37}$Ar [16] (which was produced during the 133 days on the fast neutron reactor BN600 [Beloyarskaya atomic plant, the town Zarechny, Russia]; of maximal $\nu_e$–energy up to 813 keV and $T_{1/2}$ = 35.01 days) was equal to $1.5 \cdot 10^{16}$ neutrino/s. But after 9 months the flux falls in ~300 times down to ~$5 \cdot 10^{13}$ neutrino/s.

For considered energy the cross section follows to the quadratic dependence: $\sigma_\nu \sim E_\nu^2$. So when we make a choice of neutrino source the requirements of more hard spectrum, high flux and stability become exclusively important ones. The assurance of the above mentioned conditions will give opportunities to separate the neutrino effect from background and improve significantly the reliability of experimental results.

## 2. Physical conceptions assumed as a basis for the lithium antineutrino source

It is possible avoid from serious disadvantages of reactor $\tilde{\nu}_e$–spectra (sharp decrease of $\tilde{\nu}_e$-spectrum, significant uncertainties and time instability in neutrino spectra of $^{235}$U, $^{239}$Pu, $^{238}$U, $^{241}$Pu fission products) by use of β−–decaying $^8$Li isotope with hard $\tilde{\nu}_e$-spectrum. The most simple way for creation of $^8$Li antineutrino source (lithium converter) is to arrange the lithium blanket close to the Active reactor Zone (AZ) and to ensure the effective activation: $^7$Li(n,γ)$^8$Li. The created short living $^8$Li isotope ($T_{1/2}$ = 0.84 s) emits antineutrino of well defined spectrum with maximal energy $E_{\tilde{\nu}}^{\max}$ =13.0 MeV and averaged value $\overline{E}_{\tilde{\nu}}$ = 6.5 MeV. As a result the summary $\tilde{\nu}_e$-spectrum becomes more hard compare to the reactor one (fig. 1). This type of converter realization ensures the regime of operation called as static one.

The idea of a neutrino source, based on $^8$Li decay was discussed firstly in [17] and for pulse reactors – in [18]. The questions of constructing the intensive neutrino sources with a hard spectrum, different types of lithium converters for reactors working in a stationary and pulse mode, applications of converters for neutrino researches are considered in Ref. [19 - 21]. The simple schemes of spherical construction of multi-layer converter in a static regime are presented in the fig. 4 as geometry **A** and **B**. The active zone radius $R$=23 cm corresponds to a 51.0-*liter* volume to that of the high-flux PIK reactor in Gatchina, Leningrad district [21, 22].

To compare these two types of geometry the calculations were performed (using

MAMONT code [21, 22]) for three converter thicknesses: $L_C$ = 130, 150 and 170 cm. The thickness of iron shells was 1 cm. The $D_2O$ acts as a reflector in geometry **A** and as an effective moderator in geometry **B.** The $D_2O$ thickness of $L_W$ = 30 and 15 cm are sufficient for the reflector in geometry **A** and nearly optimal for the moderator in geometry **B**. In the calculations it was assumed that one neutron with the fission spectrum escaped from the active zone per one fission. So, according to our calculations, the geometry **A** gives better results for the converter efficiency- $k$, where $k$ is equal to the number of $^8Li$ isotopes created per one neutron escaping from AZ.

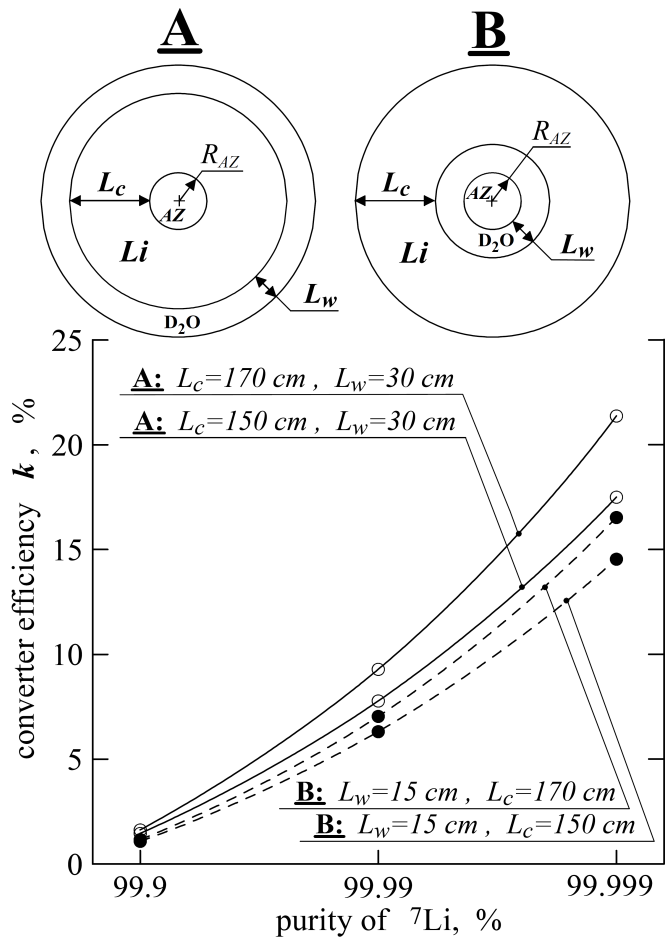

Fig. 4. Dependence of converter efficiency $k$ on $^7Li$ isotope purity (with admixture of $^6Li$) in the geometry **A** and **B** for lithium (in metallic state) thickness $L_C$, heavy water layer $L_w$ and reactor active zone radius $R_{AZ}$ = 23 cm.

The main problem is to increase the efficiency of converter and in so way to increase the hardness of the summary $\tilde{\nu}_e$-spectrum. To increase the efficiency of converter by purification of the significant mass of $^7Li$ isotope up to the 99.999% grade is highly difficulty. The constructive way may be to use $^7Li$ isotope with realistic grade of purification (about 99.99%), but in chemical compositions instead of lithium in metallic state. The perspective candidates for use as substance in a converter can be a heavy water solution of lithium hydroxides (LiOD, LiOD·$D_2O$) and lithium deuteride - LiD [23-24].

The most perspective one is the considered LiOD heavy water solution. Thus, using it permits to reduce the layer thickness $L_C$ up to ≈1 m and to reduce sharply a required mass of a high-purified lithium: for example, for $^7Li$ purification grade 99.99% in order to ensure the efficiency level $k$ = 0.075 (at LiOD concentration of 9.46 % in heavy water solution) it will be necessary the lithium mass in ~350 times less compare to the converter with $^7Li$ in metallic state only.

Comparison of efficiency $k$ (for geometry **A**) in case of different converter substances (converter with Li in metallic state, LiD, LiOD, LiOD·$D_2O$, LiOD heavy water solution of the next concentration of LiOD: 9.46%, 5.66% and 0.94%) is given in the Appendix.

In the work [25] the authors considered the alternative isotope for antineutrino converter – $B^{12}$, created by activation reaction $^{11}B(n,\gamma)^{12}B$. But it was indicated [21] that for equal isotope purification the relation of parasitic absorption on $^{10}B$ to beneficial one on $^{11}B$

$[\sigma_a(^{10}B)/\sigma_{n\gamma}(^{11}B) = 3837/0.0055]$ is considerably worse than for respective cross section on $^6Li$ and $^7Li$ $[\sigma_a(^6Li)/\sigma_{n\gamma}(^7Li) = 937/0.045]$. So, creation the neutrino source on the $^{12}B$ is highly difficulty due to technological reasons.

Significantly more hard antineutrino spectra (from active reactor zone plus decays from $^8Li$ in the lithium converter) can be ensured in the installation of the dynamic mode of operation [26 - 28]: the liquid lithium is pumped (in the close cycle) through converter and further to remote neutrino detector and back to converter in a close cycle. The propose to use (see work [23, 24]) the heavy water solution of lithium hydroxide LiOD instead of metallic lithium (as converter substance) is look the prime perspectivies as for realization of a dynamic mode the required lithium mass will be increased in about 2-4 times [26 - 28]. I.e., the conversion of metallic lithium to LiOD solution solves the problem of strong rise in price of installation and excludes the problem of safe work with metallic lithium.

### 3. Creation of antineutrino source on the base of the tandem of accelerator with the target plus lithium converter

Since the 70-th years of the XX-th century the possibility to use the spallation reaction for creation of the powerful neutrino source began attract more and more attention and were realized in the constructions of intensive neutron sources in scientific centers. Such neutron sources exist in Russia, USA, Europe, Japan and are developing: IREN, IFMIF, JSNS/J-Park (Japan), ESS, CSNS; project of electronuclear installation "Energy amplifier" by C. Rubbia, et al. [29 - 31] et al. Installations of the neutron generating targets on the accelerator (proton, electron) beam can be intensive neutron sources. In this variant for creation of intensive neutrino source we need to surround the target by lithium converter [32, 29, 30]. The large benefits of neutron source on the neutron generating target is caused by the fact that as the proton energy is increasing the neutron yield $Y_n$ (per proton) is increasing sharply: so, for proton energy $E_p = 300$ MeV the neutron yield is about $Y_n \approx (3-4)$, for $E_p = 500-600$ MeV the yield increases up to $Y_n \approx 10$; for $E_p = 1, 3, 10$ GeV the yield $Y_n$ reaches the $\approx 10, 80, 150$, respectively [33].

The used substances of neutron generating target are lead, tantalum, tungsten, uranium, mercury and beryllium (as neutron reflector and breeder). Construction of lithium (or heavy water LiOD solution) blanket around the neutron generating target will give an intensive antineutrino source.

In this work we consider the cylindrical target geometry. The target substance is tungsten isotope $^{174}W$. The scheme of lithium converter is presented on the fig.5. The $D_2O$ heavy water layer (which is effective moderator) is provided for cooling. The MCNPX code [34] and the code MAMONT (for reactor energies) [21, 28] are used for calculations. It were optimized the target size for optimization of the neutron yield for the interval $E_p = (50-300)$ МэВ. These low energies is considered with purpose to decrease the possible background in neutrino experiments and taking into account the $\pi^0$-meson production (generating the electron-photon showers) at more high energies.

For considered energies the ionization and nuclear tracks are not larger ~20 см. So, the variants of lengths $(h_t - h_h) \geq 20$ cm (fig.5) [35] for decelerating of protons are considered. The length of targets $h_t = (30-40)$ cm and channel radia $r_h = 3$ cm are discussed. The optimization was realized for purpose to enlarge neutron yield (per proton) and to minimize the neutron flux to back through the face plane of the target (the circle with the center in the point $C$ and radius $r_t$ on the fig.5). The optimization was realized in two stages – for variation of: 1) input length for beam in the interval $h_h$ =(5–20) cm; 2) radius of the target in the diapason (5–12) cm. For example for 300 MeV and total neutron yield $Y_n$ =3.61 the scattering to back fell in ~13 times up to $0.02 \cdot Y_n$. Neutron yields for optimized target (of sizes $h_t$ = 40 см, $r_t$ = 7 cm, $h_h$ = 20 cm, and $r_h$ = 3 cm) are given in the fig.6.. The obtained results are in good agreement in known experimental data and calculations for extended targets [36 - 39].

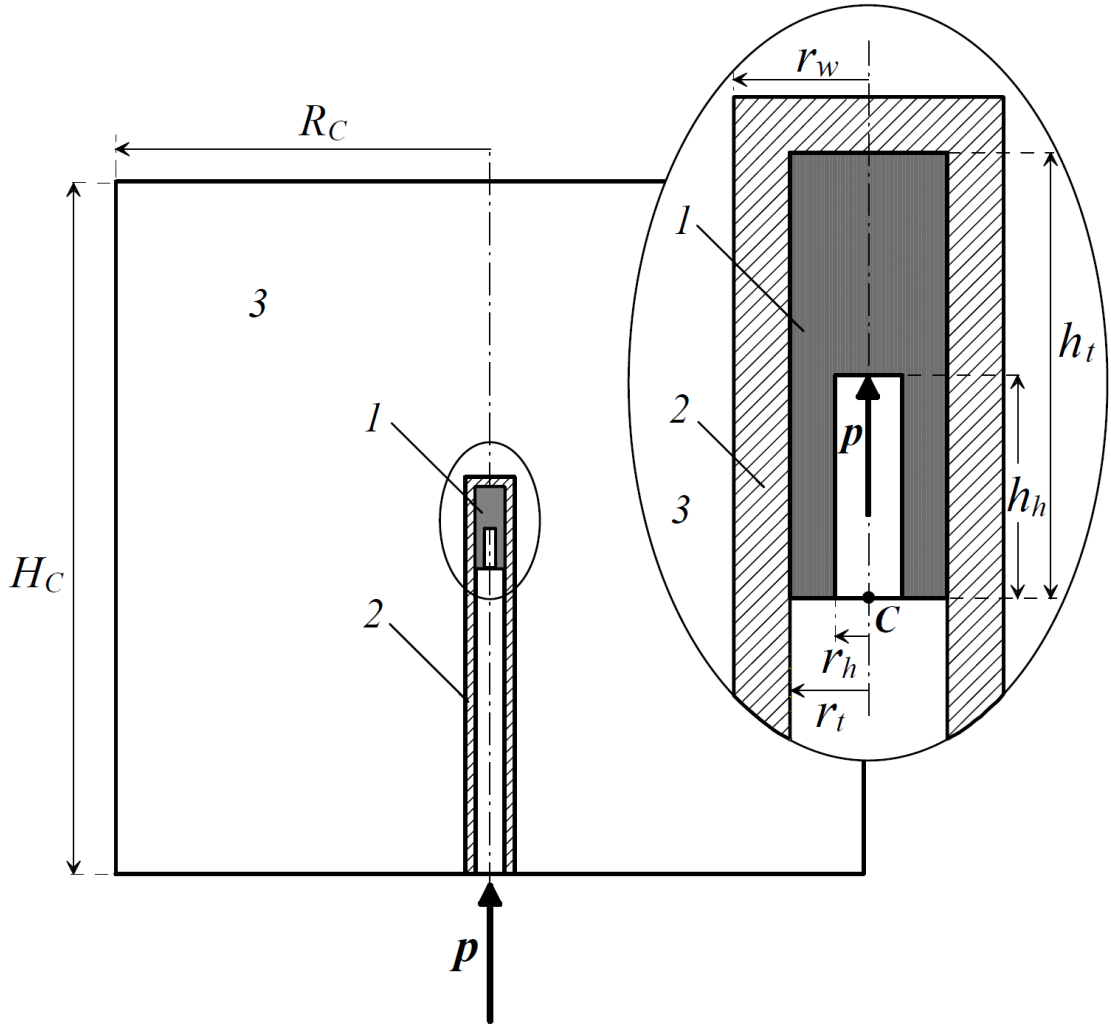

Fig. 5. The profile of the lithium converter and neutron generating target in the cylindrical geometry. *1* – tungsten (bismuth, lead) target, $h_h$ и $r_h$ – length and radia of the input channel for the proton beam; *2* – the pumping $D_2O$-channel for cooling; *3* – lithium converter.

The optimized target is cooling in the $D_2O$-channel (of thickness $(r_w - r_t) = 5$cm) and placed in the center of the cylindrical converter (see fig. 5) filled with LiOD heavy water $D_2O$ solution (of concentration - 9.46%) [23 - 24]. The cylindrical converter has the size: height $H_c$ =340 cm, radius $R_c$ = 182 cm, converter layer $L_c = (R_c - r_w) =170$ cm (as in the works [23 - 24]).

The obtained proton efficiency of the converter (number of $^8Li$ isotope, created in the converter per proton) is presented in the fig.7. Then converter antineutrino flux per solid angle $4\pi$ during time $t$ and proton current $I$ is equal to:

$$N_{\tilde{\nu}}(t) = 6.25 \cdot 10^{15} k_p(E)\, I[\text{мA}]\, t[\text{s}],$$

where: $k_p(E) = k_n(E) Y_n^{\text{eff}}(E) = k_n(E)[Y_n(E) - \delta Y_n(E)]$; $k_n$ – neutron efficiency of the converter (number of $^8$Li nuclei, created in the converter, normalized on the effective neutron yield; here $k_n \approx 0.16$); $Y_n^{\text{eff}}$ – effective neutron yield; $Y_n$ – total neutron yield; $\delta Y_n$ – correction, taking into account the loss of neutrons (mainly back scattered – to the input of the proton beam; $\delta Y_n \approx 0.02 \cdot Y_n$, see above). So, at $E_p = 300$ MeV and 1 мA of accelerator current the antineutrino flux during 1 s is $N_{\tilde{\nu}} = 3.6 \cdot 10^{15}$.

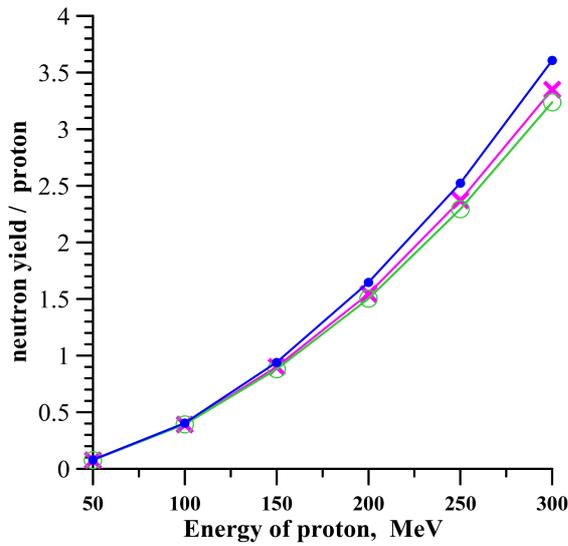 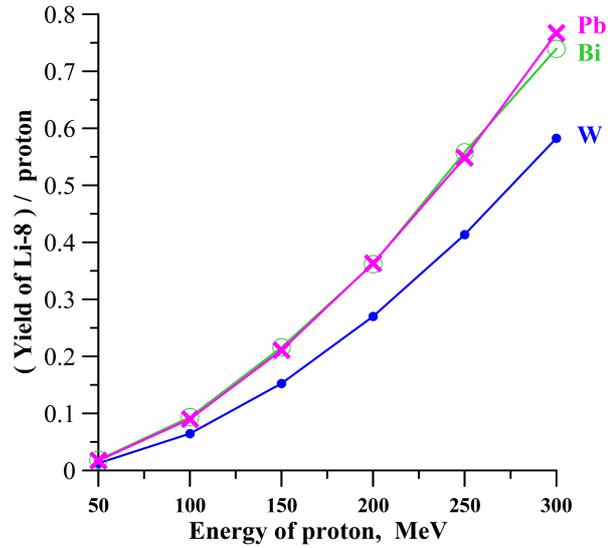

Fig. 6. Neutron yield for W, Pb and Bi-targets.   Fig. 7. Yield of $^8$Li in the converter for W, Pb and Bi-targets.

In the works [40 - 41] the authors proposed to construct $\tilde{\nu}_e$-source on the base of $^8$Li isotope. The authors note the work [29, 32] and considered the similar geometry, purity of the $^7$Li, the tandem scheme of proton accelerator (of $E_p = 60$ MeV), neutron generating $^9$Be-target and cylindrical lithium converter (height 150 cm, diameter 200 cm) filled with metallic lithium. The expected $N_{\tilde{\nu}}$-flux from the converter during 5 years (for 90% using of time and accelerator current $I = 10$ mA) is evaluated as $1.29 \cdot 10^{23}$. The authors propose to construct the accelerator (of medical type) of very large current in order to obtain high neutrino flux. Such high current will cause significant degradation of the target and necessity of large efforts for technical supports for durable operation and effective using of time. Nevertheless let us compare these data with above considered $\tilde{\nu}_e$-source with W-target, lithium converter (filled with LiOD heavy water solution of concentration -9.46%), 90% time using and proton current - $I = 10$ мA. Then for proton energy $E_p = 300$ и 100 MeV the fluxes $1.29 \cdot 10^{23}$ $\tilde{\nu}_e$ will be obtained in 46 and 411 days respectively.

It is necessary to compare neutron yields in case different targets. Calculations were realized for lead, bismuth and tungsten ($^{174}$W) targets. The all three targets ensure close values for neutron yields (per proton) – see fig.6. The difference of maximal yield ($^{174}$W target) and minimal one (Bi target) is not larger than ~ 10%. But in case of converter (i.e., neutron generating target is placed inside the converter (filled with LiOD heavy water solution)) we have an inverse "picture": the proton converter efficiency (number of $^8$Li isotopes created in the converter per proton) depending on proton energy $k_p(E)$ is maximal for Pb-target and minimal for W-target (fig.7). The cause of obtained inversion is the converter itself: for calculations of neutron yields $Y_n$ the used geometry is "ideal" - the target is "placed" in the vacuum. But in case of the real geometry the neutrons emitted from the target enter to the converter and can be scattered back to the target and be absorbed. Also it is possible the scattering back to the input channel of the proton beam. The fig.8 illustrates the neutron balance on the boundary of the W- and Pb-target. The vertical axe is the directional current through the surface: the positive values respect to the current of neutrons escaped from the target; negative values respect the current for neutrons scattered inside the target. The algebraic sum of escaped neutrons and neutrons scattered inside the target gives the number of neutrons captured out of the target. So, for Pb-target the number of events for capture out of the Pb-target (1.397) is larger compare to the similar events for W-target (1.008) and this is cause of the observed inversion phenomena.

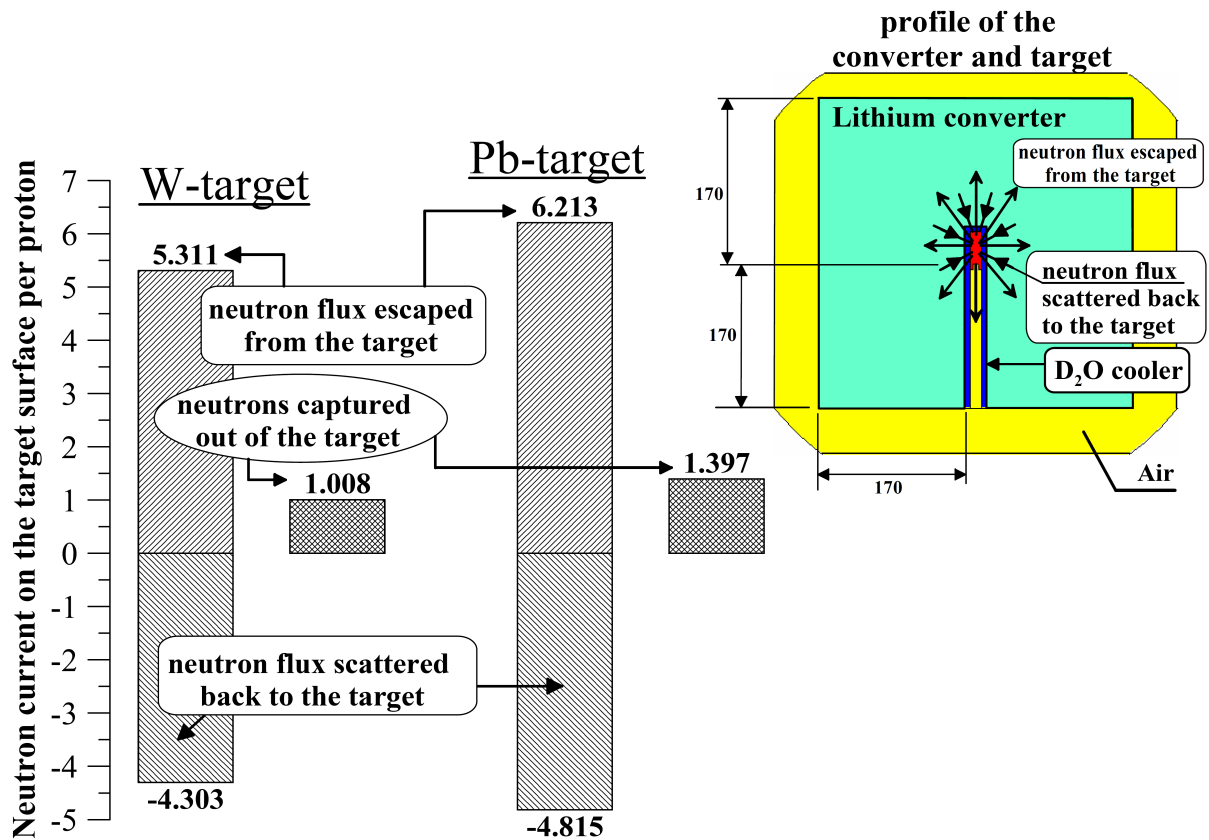

Fig. 8. Neutron balance on the boundary of the W- and Pb-target.

It is important to note that realization of the project [40 - 41] with lithium in metallic state will request ~ 2.7 t of lithium of purification 99.99% on the $^7$Li isotope compare to ~ 1.1 t for the discussed here antineutrino source on the base of converter with heavy water LiOD solution (concentration - 9.46%) and W-target. I.e., using of heavy water LiOD solution will allow to decrease strongly the price of creation for expensive physical installation.

It will be helpful to compare $^8$Li yield in the same geometry for converter filled with: 1) lithium in metallic state; 2) heavy water solution of LiOD. We propose to compare $^8$Li yield for (above mentioned) energy interval of protons $E_p$ =(50–300) МэВ and different converter substances (metallic lithium and heavy water solution of LiOD concentration 9.46%). In discussed cases the neutron fields in the converter will be vary widely. So the most universal geometry for the converter will be spherical one. Let us consider the spherical converter geometry with thickness of converter substance about 1 m (that is close to [40 - 41]). For these comparative calculations we will use the cylindrical tungsten ($^{174}$W) target with input for proton beam and the same optimized sizes as in the fig.5. The geometry of the converter with target is presented in the fig.9. We emphasize that it is the idealized geometry of close spherical converter (with target in the center) for calculation of the maximal $^8$Li yield.

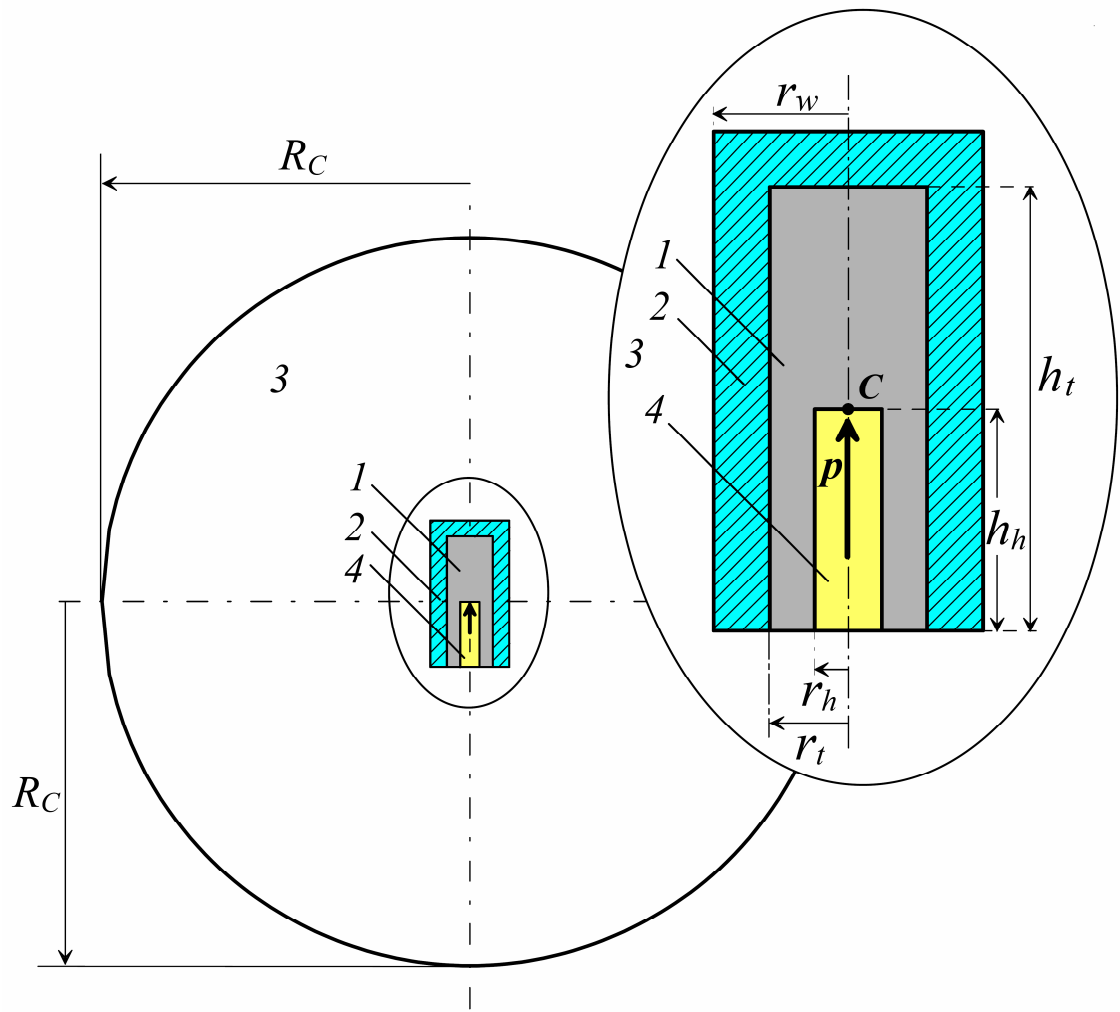

Fig. 9. The profile of the lithium converter in the spherical geometry of radius $R_c$ = 112 cm and center in the point C (see enlarged profile for the target): *1* – tungsten target; $h_t$ and $r_t$ – length and radius of the target; $h_h$ and $r_h$ – length and radius of the input channel for the proton beam; *2* – the D$_2$O-channel for cooling of radia $r_w$; *3* – lithium converter; *4* – empty (filled with air) input channel. The sizes of the target, radia $r_w$ and thickness of D$_2$O-channel are the same as in the fig. 5.

The dependence of $^8$Li yield from the proton energy is given in the fig. 10. For the thickness about 1 m= ($R_c$ - $r_w$) the $^8$Li yield for LiOD heavy water solution is larger in (3.2 - 3.4) times compare to the lithium in metallic state. The obtained results for LiOD solution respect to the pure deuterium. But in fact some parasitic admixture of light hydrogen $_1$H$^1$ always exists in heavy water D$_2$O. We considered several grades of deuterium purification (in %) in heavy water solution of LiOD. It were obtained that realizable purification (about $P_D$ = 99.0 and 98.0%) does not distort strongly the $^8$Li yield: compare to converter with lithium in metallic state the creation of $^8$Li for these grades of D-purification is larger more than in 2.7 and 2.4 times.

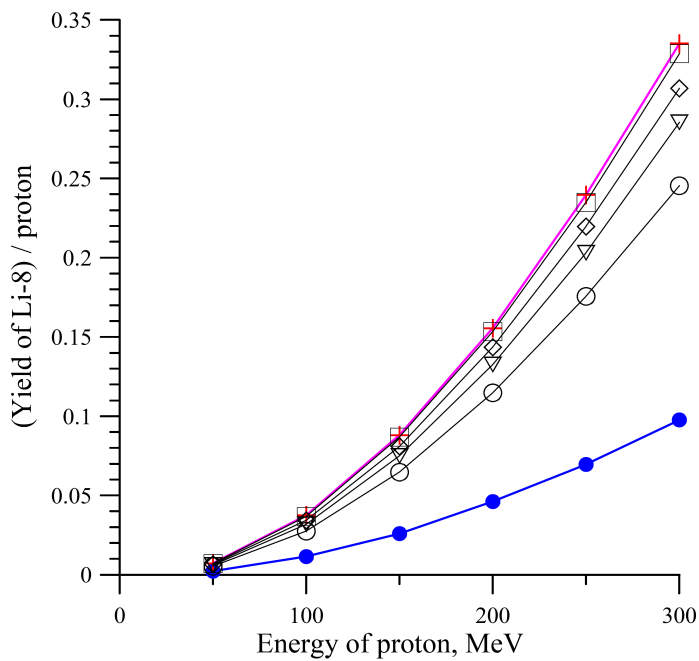

Fig. 10. Yield of $^8$Li in the spherical geometry for converter filled with lithium in metallic state and heavy water solution of LiOD (of different grade of deuterium purification). ● - yield of $^8$Li for lithium in metallic state. Yield of $^8$Li for heavy water solution of LiOD with grade of deuterium purification $P_D$ : + - 100%; □ - 99.9%; ◇ - 99.5%; ∇ - 99.0%; ○ - 98.0%.

Use of heavy water LiOD solution allows to moderate neutrons more effectively than in Li case. So it is possible: 1) to decrease significantly the requested volume of the converter and 2) to decrease strongly the required mass of high purified $^7$Li. (for discussed geometry the mass of lithium in metallic state is ~3380 kg compare to ~190 kg in LiOD heavy water solution). The data for yield of $^8$Li are also presented in the table 1.

Table 1.
Yields of $^8$Li isotope per proton (of the energy $E_p$) in the lithium converter (for spherical converter geometry) in cases of: lithium in metallic state (Li); heavy water solution for 9.46% of LiOD with different deuterium purity $P_D$.

| $E_p$, MeV | Yields of $^8$Li for metallic lithium | Yields of $^8$Li for 9.46% heavy water solution of LiOD | | | | |
|---|---|---|---|---|---|---|
| | | $P_D$=100% | $P_D$=99.9% | $P_D$=99.5% | $P_D$=99.0% | $P_D$=98.0% |
| 50 | 2.22E-3 | 7.20E-3 | 7.03E-3 | 6.62E-3 | 6.07E-3 | 5.29E-3 |
| 100 | 1.14E-2 | 3.74E-2 | 3.68E-2 | 3.43E-2 | 3.18E-2 | 2.75E-2 |
| 150 | 2.59E-2 | 8.81E-2 | 8.69E-2 | 8.09E-2 | 7.49E-2 | 6.48E-2 |
| 200 | 4.62E-2 | 1.56E-1 | 1.53E-1 | 1.44E-1 | 1.33E-1 | 1.15E-1 |
| 250 | 6.95E-2 | 2.40E-1 | 2.35E-1 | 2.20E-1 | 2.03E-1 | 1.76E-1 |
| 300 | 9.77E-2 | 3.35E-1 | 3.29E-1 | 3.07E-1 | 2.85E-1 | 2.45E-1 |

## 4. Conclusion

It was considered two variants of intensive $\tilde{\nu}_e$-source on the base of $^7$Li isotope (with purity 99.99%) basing on the 1) nuclear reactor (as neutron source), and 2) proton accelerator and neutron generating target (tungsten, bismuth and lead). Owing to the hardness and well defined $\tilde{\nu}_e$-spectrum the $^8$Li isotope becomes a very perspective antineutrino source.

The reactor realization of intensive $\tilde{\nu}_e$-source ensures the summary $\tilde{\nu}_e$-spectrum of β⁻-decaying nuclei (fission fragments) plus spectrum of $^8$Li. This leads to creation of more hard spectrum [21, 24, 28, 32, 42, 43] and allows to increase in times the cross section on protons and deuterons compare to the results in the pure reactor neutrino spectrum (which have significant uncertainties (~ 6%).

Realization of the accelerator variant of the $\tilde{\nu}_e$-source will ensure the pure lithium $\tilde{\nu}_e$-spectrum. It was discussed the variants of the tandem of proton accelerators with energy $E_p$ =50-300 MeV and tungsten, bismuth and lead targets plus lithium converter (filled with heavy water solution of LiOD). It were calculated the neutron yields from W-, Bi-, Pb-targets, creation of $^8$Li isotope and expected $\tilde{\nu}_e$-fluxes. The considered $\tilde{\nu}_e$-source (in cylindrical geometry – fig. 5) will ensure: 1) to decrease strongly the required mass of high purified $^7$Li isotope (in 2.4 times compare to [40, 41] and will allow to decrease very significantly the cost of the installation) in the installation for neutrino investigations and 2) the $1.3 \cdot 10^{22}$ $\tilde{\nu}_e$-flux for proton energy $E_p$ =300 and 100 MeV (at the current 1 mA) in 46 and 411 days, respectively.

The more perspective will be $\tilde{\nu}_e$-source of smaller size (about 1 m in radius for spherical geometry or for cylindrical geometry with equivalent dimensions). It was demonstrated that yield of $^8$Li for LiOD solution (for concentration 9.46% and deuterium purification 99%) is larger in 2.7-2.9 times compare to the lithium in metallic state. In addition use of LiOD solution will request the pure lithium in 17.8 times less compare to the converter filled with metallic lithium.

The proposed intensive antineutrino source (lithium converter) will ensure hard and well determined spectrum (that extremely important for decrease of errors) and (owing to proposed of lithium chemical compounds) can be constructed as compact installation that is necessary for investigations of neutrino oscillation.

The work was partially supported by the Russian Foundation for Basic Research Grants no. 13-02-12106 ofi-m, 14-22-03040 (part I) ofi-m and Swiss National Science Foundation grant no IZ73Z0_152485 SCOPES.

## Appendix

Converter efficiency $k$ (top value in the cells) and mass of lithium $m$ (lower value in brackets in the cells) for converter thickness $L_C$ (in geometry **A**) in case of different converter substances (Li - lithium in metallic state; LiD; LiOD; LiOD·$D_2O$; LiOD heavy water solution of three concentrations of LiOD) [23, 29]. In this table the data for lithium in metallic state were recalculated again according to specified density of $^7$Li isotope - 0.553 g/cm$^3$ [44].

| $L_C$, cm | $k$, %, ($m$, kg) | | | | | | |
|---|---|---|---|---|---|---|---|
| | Li | LiD | LiOD | LiOD·D2O | Heavy water solution of LiOD for the concentrations: | | |
| | | | | | 9,46% | 5,66% | 0,94% |
| 10 | 0,3 (59,0) | 2,0 (66,3) | 1,5 (44,7) | 3,3 (32,7) | 1,1 (3,5) | 0,9 (2.0) | 0,3 (0,3) |
| 20 | 0,7 (165,3) | 5,8 (185,8) | 4,4 (125,3) | 7,7 (91,5) | 3,7 (9,7) | 2,6 (5,6) | 1,0 (0,9) |
| 30 | 1,2 (332,7) | 9,4 (374,0) | 8,0 (252,2) | 10,8 (184,1) | 6,2 (19,5) | 5,0 (11,3) | 1,7 (1,8) |
| 40 | 1,6 (575,2) | 12,2 (646,5) | 10,2 (436,0) | 11,9 (318,3) | 7,7 (33,7) | 6,5 (19,5) | 2,3 (3,1) |
| 50 | 2,0 (906,6) | 14,0 (1019,0) | 11,6 (687,2) | 12,2 (501,7) | 8,9 (53,2) | 7,6 (30,7) | 2,7 (4,8) |
| 70 | 2,8 (1891,9) | 15,5 (2126,5) | 13,2 (1434,0) | 12,6 (1047,0) | 10,2 (111,0) | 8,9 (64,1) | 3,4 (10,1) |
| 90 | 3,8 (3399,8) | 15,8 (3821,3) | 13,5 (2576,9) | 12,65 (1881,5) | 11,0 (199,5) | 9,6 (115,1) | 3,9 (18,2) |
| 110 | 5,0 (5541,5) | (6228,5) | (4200,1) | (3066,8) | 11,5 (325,2) | 10,2 (187,6) | 4,2 (29,7) |
| 130 | 6,4 (8428,1) | (9473,0) | (6388,0) | (4664,26) | 11,8 (494,6) | 10,5 (285,4) | 4,5 (45,1) |
| 150 | 7,8 (12170,8) | (13679,8) | (9224,8) | (6735,6) | 12,0 (714,2) | 10,8 (412,1) | 4,7 (65,1) |